# Strongly Enhanced Gilbert Damping in 3*d* Transition Metal Ferromagnet Monolayers in Contact with Topological Insulator Bi$_2$Se$_3$

Y. S. Hou[1], and R. Q. Wu[1]

[1] Department of Physics and Astronomy, University of California, Irvine, California 92697-4575, USA

## Abstract

Engineering Gilbert damping of ferromagnetic metal films is of great importance to exploit and design spintronic devices that are operated with an ultrahigh speed. Based on scattering theory of Gilbert damping, we extend the torque method originally used in studies of magnetocrystalline anisotropy to theoretically determine Gilbert dampings of ferromagnetic metals. This method is utilized to investigate Gilbert dampings of 3*d* transition metal ferromagnet iron, cobalt and nickel monolayers that are contacted by the prototypical topological insulator Bi$_2$Se$_3$. Amazingly, we find that their Gilbert dampings are strongly enhanced by about one order in magnitude, compared with dampings of their bulks and free-standing monolayers, owing to the strong spin-orbit coupling of Bi$_2$Se$_3$. Our work provides an attractive route to tailoring Gilbert damping of ferromagnetic metallic films by putting them in contact with topological insulators.

Email: wur@uci.edu

# I. INTRODUCTION

In ferromagnets, the time-evolution of their magnetization **M** can be described by the Landau-Lifshitz-Gilbert (LLG) equation [1-3]

$$\frac{d\mathbf{M}}{dt} = -\gamma \mathbf{M} \times \mathbf{H}_{eff} + \frac{\alpha}{M_S} \mathbf{M} \times \frac{d\mathbf{M}}{dt} \qquad (1),$$

where $\gamma = g\mu_0\mu_B/\hbar$ is the gyromagnetic ratio, and $M_S = |\mathbf{M}|$ is the saturation magnetization. The first term describes the precession motion of magnetization **M** about the effective magnetic field, $\mathbf{H}_{eff}$, which includes contributions from external field, magnetic anisotropy, exchange, dipole-dipole and Dzyaloshinskii-Moriya interactions [3]. The second term represents the decay of magnetization precession with a dimensionless parameter $\alpha$, known as the Gilbert damping [4-8]. Gilbert damping is known to be important for the performance of various spintronic devices such as hard drives, magnetic random access memories, spin filters, and magnetic sensors [3, 9, 10]. For example, Gilbert damping in the free layer of reader head in a magnetic hard drive determines its response speed and signal-to-noise ratio [11, 12]. The bandwidth, insertion loss, and response time of a magnetic thin film microwave device also critically depend on the value of $\alpha$ in the film [13].

The rapid development of spintronic technologies calls for the ability of tuning Gilbert damping in a wide range. Several approaches have been proposed for the engineering of Gilbert damping in ferromagnetic (FM) thin films, by using non-magnetic or rare earth dopants, adding different seed layers for growth, or adjusting composition ratios in the case of alloy films [9, 14-16]. In particular, tuning $\alpha$ via contact with other materials such as heavy metals, topological insulators (TIs), van der Waals monolayers or magnetic insulators is promising as the selection of material combinations is essentially unlimited. Some of these materials may have fundamentally different damping mechanism and offer opportunity for studies of new phenomena such as spin-orbit torque, spin-charge conversion, and thermal-spin-behavior[17, 18].

In this work, we systematically investigate the effect of $Bi_2Se_3$ (BS), a prototypical TI, on the Gilbert damping of 3$d$ transition metal (TM) Fe, Co and Ni monolayers (MLs) as they are in contacted with each other. We find that the Gilbert dampings in the TM/TI combinations are enhanced by about an order of magnitude than their counterparts in bulk Fe, Co and Ni as well as in the free-standing TM MLs. This drastic enhancement can be attributed to the strong spin-orbit coupling (SOC) of the TI substrate and might also be related to its topological nature. Our work introduces an appealing way to engineer Gilbert dampings of FM metal films by using the peculiar physical properties of TIs.

## II. COMPUTATIONAL DETAILS

Our density functional theory (DFT) calculations are carried out using the Vienna *Ab-initio* Simulation Package (VASP) at the level of the generalized gradient approximation [19-22]. We treat Bi-6s6p, Se-4s4p, Fe-3d4s, Co-3d4s and Ni-3d4s as valence electrons and employ the projector-augmented wave pseudopotentials to describe core-valence interactions [23, 24]. The energy cutoff of plane-wave expansion is 450 eV [22]. The BS substrate is simulated by five quintuple layers (QLs), with an in-plane lattice constant of $a_{BS}$ = 4.164 Å and a vacuum space of 13 Å between slabs along the normal axis. For the computational convenience, we put Fe, Co and Ni MLs on both sides of the BS slab. For the structural optimization of the BS/TM slabs, a 6×6×1 Gamma-centered k-point grid is used, and the positions of all atoms except those of the three central BS QLs are fully relaxed with a criterion that the force on each atom is less than 0.01 eV/Å. The van der Waals (vdW) correction in the form of the nonlocal vdW functional (optB86b-vdW) [25, 26] is included in all calculations.

The Gilbert dampings are determined by extending the torque method that we developed for the study of magnetocrystalline anisotropy [27, 28]. To ensure the numerical convergence, we use very dense Gamma-centered k-point grids and, furthermore, large numbers of unoccupied bands. For example, the first Brillouin zone of BS/Fe is sampled by a 37×37×1 Gamma-centered k-point grid, and the number of bands for the second-variation step is set to 396, twice of the number (188) of the total valence electrons. More computational details are given in Appendix A. Magnetocrystalline anisotropy energies are determined by computing total energies with different magnetic orientations [29].

## III. TORQUE METHOD OF DETERMINING GILBERT DAMPING

According to the scattering theory of Gilbert damping [30, 31], the energy dissipation rate of the electronic system with a Hamiltonian, $H(t)$, is determined by

$$\dot{E}_{dis} = -\pi\hbar \sum_{ij}\sum_{\mu\nu} \dot{u}_\mu \dot{u}_\nu \langle\psi_i|\frac{\partial H}{\partial u_\mu}|\psi_j\rangle\langle\psi_j|\frac{\partial H}{\partial u_\nu}|\psi_i\rangle \delta(E_F - E_i)\delta(E_F - E_j) \quad (2).$$

Here, $E_F$ is the Fermi level and $u$ is the deviation of normalized magnetic moment away from its equilibrium, i.e., $m = m_0 + u$ with $m_0 = M_0/M_s$. On the other hand, the time derivative of the magnetic energy in the LLG equation is [32]

$$\dot{E}_{mag} = \mathbf{H}_{eff} \cdot \frac{d\mathbf{M}}{dt} = \frac{M_S}{\gamma}\sum_{\mu\nu}\alpha_{\mu\nu}\dot{m}_\mu \dot{m}_\nu \quad (3).$$

By taking $\dot{E}_{dis} = \dot{E}_{mag}$, one obtains the Gilbert damping as:

$$\alpha_{\mu\nu} = -\frac{\pi\hbar\gamma}{M_S}\sum_{ij}\langle\psi_i|\frac{\partial H}{\partial u_\mu}|\psi_j\rangle\langle\psi_j|\frac{\partial H}{\partial u_\nu}|\psi_i\rangle \delta(E_F - E_i)\delta(E_F - E_j) \quad (4).$$

Note that, to obtain Eq. (4), we use $\partial m = \partial u$ since the equilibrium normalized magnetization $m_0$ is a constant. In practical numerical calculations, $\delta(E_F - E)$ is typically substituted by the Lorentzian function $L(\varepsilon) = 0.5\Gamma / \left[\pi(\varepsilon - \varepsilon_0)^2 + \pi(0.5\Gamma)^2\right]$. The half maximum parameter, $\Gamma = 1/\tau$, is adjusted to reflect different scattering rates of electron-hole pairs created by the precession of magnetization **M** [10]. This procedure has been already used in several *ab initio* calculations for Gilbert dampings of metallic systems [8, 9, 32-35], where the electronic responses play the major role for energy dissipation.

In this work, we focus on the primary Gilbert damping in FM metals that arises from SOC [10, 36-38]. There are two important effects in a uniform precession of magnetization **M**, when SOC is taken into consideration. The first is the Fermi surface breathing as **M** rotates, i.e., some occupied states shift to above the Fermi level and some unoccupied states shift to below the Fermi level. The second is the transition between different states across the Fermi level as the precession can be viewed as a perturbation to the system. These two effects generate electron-hole pairs near the Fermi level and their relaxation through lattice scattering leads to the Gilbert damping.

Now we demonstrate how to obtain the Gilbert damping due to SOC through extending our previous torque method [27]. The general Hamiltonian in Eq. (4) can be replaced by $\boldsymbol{H}_{SOC} = \sum_j \xi(r_j) \boldsymbol{l}_j \cdot \boldsymbol{s}$ [4, 27] where the index $j$ refers to atoms, and $\boldsymbol{l}_j = -i\boldsymbol{r}_j \times \nabla$ and $\boldsymbol{s}$ are orbital and spin operators, respectively. This is in the same spirit for the determination of the magnetocrystalline anisotropy [27], for which our torque method is recognized as a powerful tool in the framework of spin-density theory [27]. When $\boldsymbol{m}$ points at the direction of $\boldsymbol{n}(\theta,\varphi) = (m_x, m_y, m_z)$, the term $\boldsymbol{l} \cdot \boldsymbol{s}$ in $\boldsymbol{H}_{SOC}$ is written as follows:

$$\boldsymbol{l} \cdot \boldsymbol{s} = s_n \left( l_z \cos\theta + \frac{1}{2} l_+ e^{-i\varphi} \sin\theta + \frac{1}{2} l_- e^{i\varphi} \sin\theta \right) +$$
$$\frac{1}{2} s_+ \left( -l_z \sin\theta - l_+ e^{-i\varphi} \sin^2 \frac{\theta}{2} + l_- e^{i\varphi} \cos^2 \frac{\theta}{2} \right) + \quad (5)$$
$$\frac{1}{2} s_- \left( -l_z \sin\theta + l_+ e^{-i\varphi} \cos^2 \frac{\theta}{2} - l_- e^{i\varphi} \sin^2 \frac{\theta}{2} \right)$$

To obtain the derivatives of $\boldsymbol{H}$ in Eq. (4), we assume that the magnitude of $\boldsymbol{M}$ is constant as its direction changes [36]. The processes of getting angular derivatives of $\boldsymbol{H}$ are straightforward and the results are given by Eq. (A1)-(A5) in Appendix B.

## IV. RESULTS AND DISCUSSION

In this section, we first show that our approach of determining Gilbert damping works well for FM metals such as 3$d$ TM Fe, Co and Ni bulks. Following that, we demonstrate the strongly enhanced Gilbert dampings of Fe, Co and Ni MLs due to the contact with BS and then discuss the underlying physical mechanism of these enhancements.

### A. Gilbert dampings of 3$d$ TM Fe, Co and Ni bulks

Gilbert dampings of 3$d$ TM bcc Fe, hcp Co and fcc Ni bulks calculated by means of our extended torque method are consistent with previous theoretical results [10]. As shown in Fig. 1, the intraband contributions decrease whereas the interband contributions increase as the scattering rate $\Gamma$ increases. The minimum values of $\alpha$ have the same magnitude as those in Ref. [10] for all three metals, showing the applicability of our approach for the determination of Gilbert dampings of FM metals.

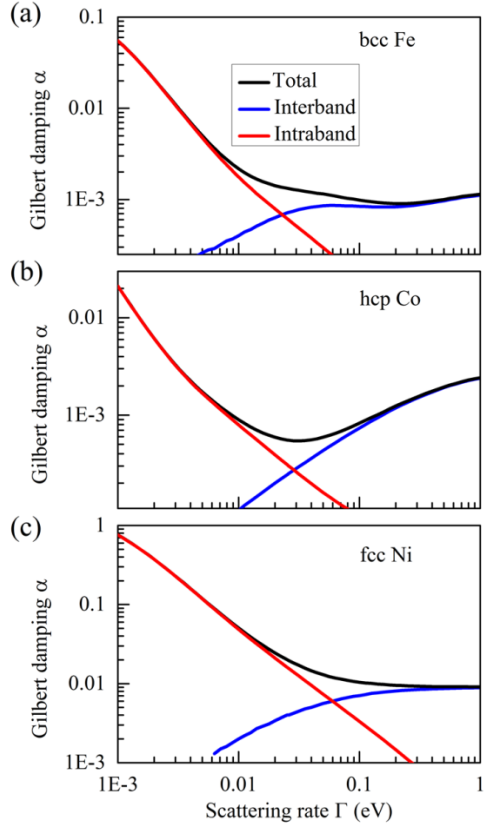

*Figure 1 (color online) Gilbert dampings of (a) bcc Fe, (b) hcp Co and (c) fcc Ni bulks. Black curves give the total Gilbert damping. Red and blue curves give the intraband and interband contributions to the total Gilbert damping, respectively.*

**B. Strongly enhanced Gilbert dampings of Fe, Co and Ni MLs in contact with BS**

We now investigate the magnetic properties of heterostructures of BS and Fe, Co and Ni MLs. BS/Fe is taken as an example and its atom arrangement is shown in Fig. 2a. From the spatial distribution of charge density difference $\Delta\rho = \rho_{BS+Fe-ML} - \rho_{BS} - \rho_{Fe-ML}$ in Fig. 2b, we see that there is fairly obvious charge transfer between Fe and the topmost Se atoms. By taking the average of $\Delta\rho$ in the ***xy*** plane, we find that charge transfer mainly takes place near the interface (Fig.2c). Furthermore, the charge transfer induces non-negligible magnetization in the topmost QL of BS (Fig. 2b). Similar charge transfers and induced magnetization are also found in BS/Co and BS/Ni (Fig. A1 and Fig. A2 in Appendix C). These suggest that interfacial interactions between BS and 3d TMs are very strong. Note that BS/Fe and BS/Co have in-plane easy axes whereas the BS/Ni has an out-of-plane one.

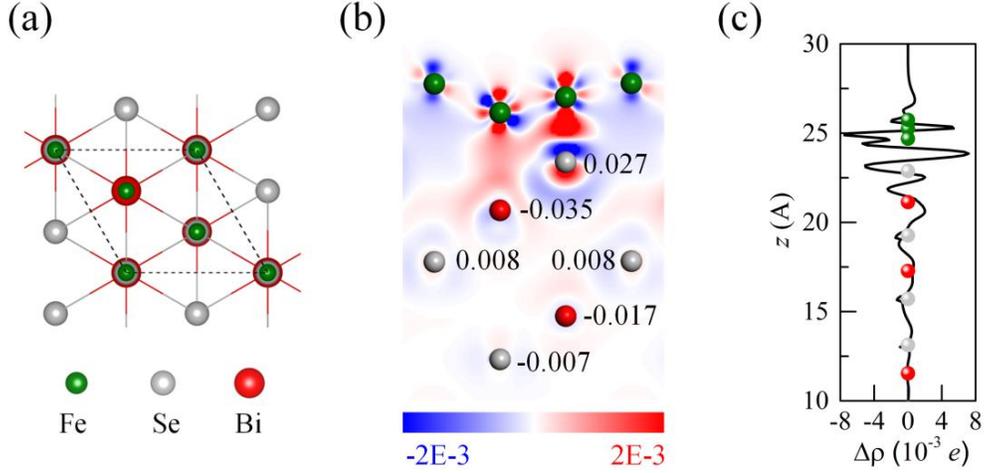

Figure 2 (color online) (a) Top view of atom arrangement in BS/Fe. (b) Charge density difference $\Delta\rho$ near the interface in BS/Fe. Numbers give the induced magnetic moments (in units of $\mu_B$) in the topmost QL BS. Color bar indicates the weight of negative (blue) and positive (red) charge density differences. (c) Planer-averaged charge density difference $\Delta\rho$ in BS/Fe. In (a), (b), (c), dark green, light gray and red balls represent Fe, Se and Bi atoms, respectively.

Fig. 3a and 3b show the $\Gamma$ dependent Gilbert dampings of BS/Fe, BS/Co and BS/Ni. It is striking that Gilbert dampings of BS/Fe, BS/Co and BS/Ni are enhanced by about one or two orders in magnitude from the counterparts of Fe, Co and Ni bulks as well as their free-standing MLs, depending on the choice of scattering rate in the range from 0.001 to 1.0 eV. Similar to Fe, Co and Ni bulks, the intraband contributions monotonically decrease while the interband contributions increase as the scattering rate $\Gamma$ gets larger (Fig. A3 in Appendix D). Note that our calculations indicate that there is no obvious difference between the Gilbert dampings of BS/Fe when five- and six-QL BS slabs are used (Fig. A4 in Appendix E). This is consistent with the experimental observation that the interaction between the top and bottom topological surface states is negligible in BS thicker than five QLs [39].

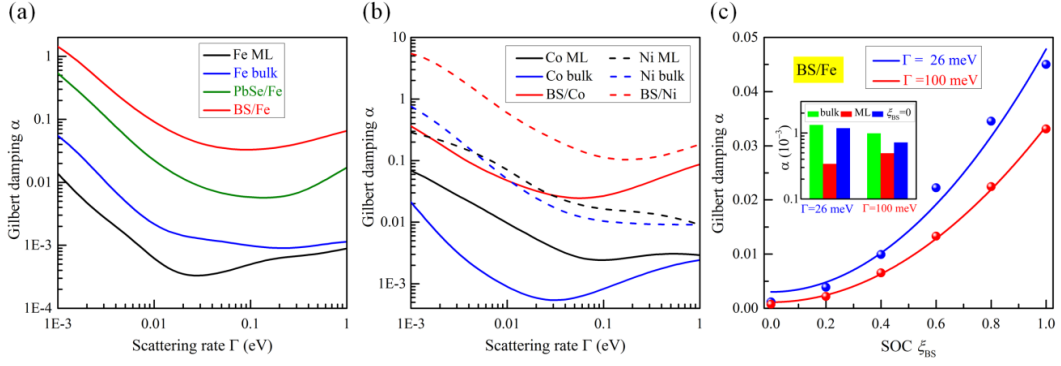

*Figure 3 (color online) Scattering rate $\Gamma$ dependent Gilbert dampings of (a) Fe ML, bcc Fe bulk, BS/Fe and PbSe/Fe, (b) Co ML, hcp Co bulk, BS/Co, Ni ML, fcc Ni bulk and BS/Ni. (c) Dependence of the Gilbert damping of BS/Fe on the scaled SOC $\xi_{BS}$ of BS in the range from zero ($\xi_{BS}=0$) to full strength ($\xi_{BS}=1$). Solid lines show the fitting of Gilbert damping $\alpha_{BS/Fe}$ to Eq. (6). The inset shows Gilbert damping comparisons between BS/Fe at $\xi_{BS}=0$, bcc Fe bulk and Fe ML.*

As is well-known, TIs are characterized by their strong SOC and topologically nontrivial surface states. An important issue is how they affect the Gilbert dampings in BS/TM systems. Using BS/Fe as an example, we artificially tune the SOC parameter $\xi_{BS}$ of BS from zero to full strength and fit the Gilbert damping $\alpha_{BS/Fe}$ in powers of $\xi_{BS}$ as

$$\alpha_{BS/Fe} = \alpha_2 \xi_{BS}^2 + \alpha_{BS/Fe}(\xi_{BS}=0) \qquad (6).$$

As shown in Fig. 3c, we obtain two interesting results: (I) when $\xi_{BS}$ is zero, the calculated residual Gilbert damping $\alpha_{BS/Fe}(\xi_{BS}=0)$ is comparable to Gilbert dampings of bcc Fe bulk and Fe free-standing ML (see the inset in Fig. 3c); (II) Gilbert damping $\alpha_{BS/Fe}$ increases almost linearly with $\xi_{BS}^2$, similar to previous results [36]. These reveal that the strong SOC of BS is crucial for the enhancement of Gilbert damping.

To gain insight into how the strong SOC of BS affects the damping of BS/Fe, we explore the *k*-dependent contributions to Gilbert damping, $\alpha_{BS/Fe}$. As shown in Fig. 4a, many bands near the Fermi level show strong intermixing between Fe and BS orbitals (marked by black arrows). Accordingly, these *k*-points have large contributions to the Gilbert damping (marked by red arrows in Fig. 4b). However, if the hybridized states are far away from the Fermi level, they make almost zero contribution to the Gilbert damping. Therefore, we conclude that only hybridizations at or close to Fermi level have dominant influence on the Gilbert damping. This is understandable, since energy differences $E_F$-$E_i$ and $E_F$-$E_j$ are important in the Lorentzian functions in Eq. (4).

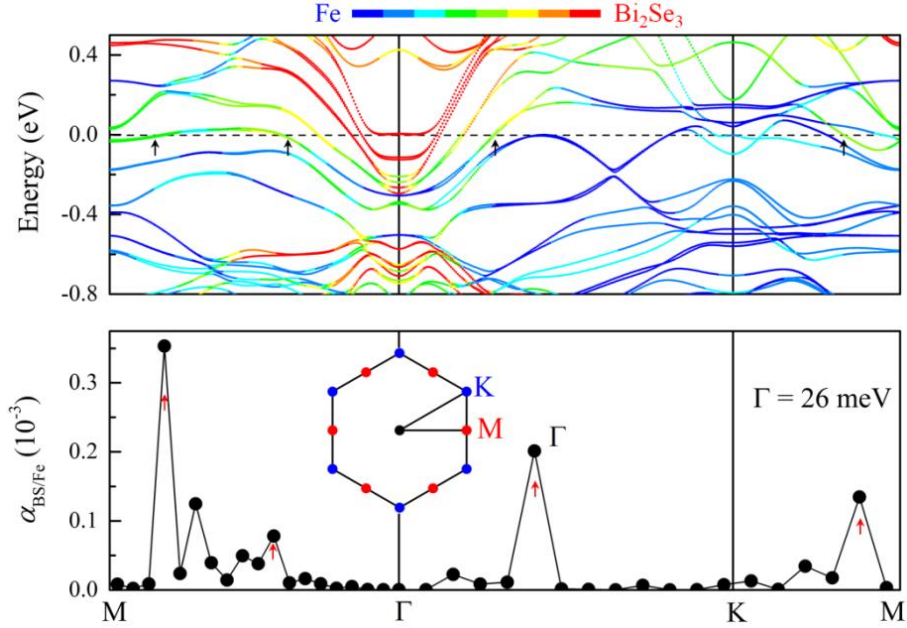

*Figure 4 (color online) (a) DFT+SOC calculated band structure of BS/Fe. Color bar indicates the weights of BS (red) and Fe ML (blue). Black dashed line indicates the Fermi level. (b) k-dependent contributions to Gilbert damping $\alpha_{\text{BS/Fe}}$ at scattering rate $\Gamma = 26\,\text{meV}$. Inset shows the first Brillouin zone and high-symmetry k-points $\Gamma$, $K$ and $M$.*

It appears that there is no direct link between the topological nature of BS and the strong enhancement of Gilbert damping. The main contributions to Gilbert damping are not from the vicinity around the $\Gamma$-point, where the topological nature of BS manifests. Besides, BS should undergo a topological phase transition from trivial to topological as its SOC $\xi_{\text{BS}}$ increases [40]. If the topological nature of BS dictates the enhancement of Gilbert damping, one should expect a kink in the $\alpha(\xi_{\text{BS}})$ curve at this phase transition point but this is obviously absent in Fig. 3c.

To dig deeper into this interesting issue, we replace the topologically nontrivial BS with a topologically trivial insulator PbSe, because the latter has a nearly the same SOC as the former. As shown in Fig. 3a, the Gilbert damping of PbSe/Fe is noticeably smaller than that of BS/Fe, although both are significantly enhanced from the values of $\alpha$ of Fe bulk and Fe free-standing ML. Taking the similar SOC and surface geometry between BS and PbSe (Fig. A5 in Appendix F), the large difference between the Gilbert dampings of BS/Fe and PbSe/Fe suggests that the topological nature of BS still has an influence on Gilbert damping. One possibility is that the BS surface is metallic with the presence of the time-reversal protected topological surface states and hence the interfacial hybridization is stronger.

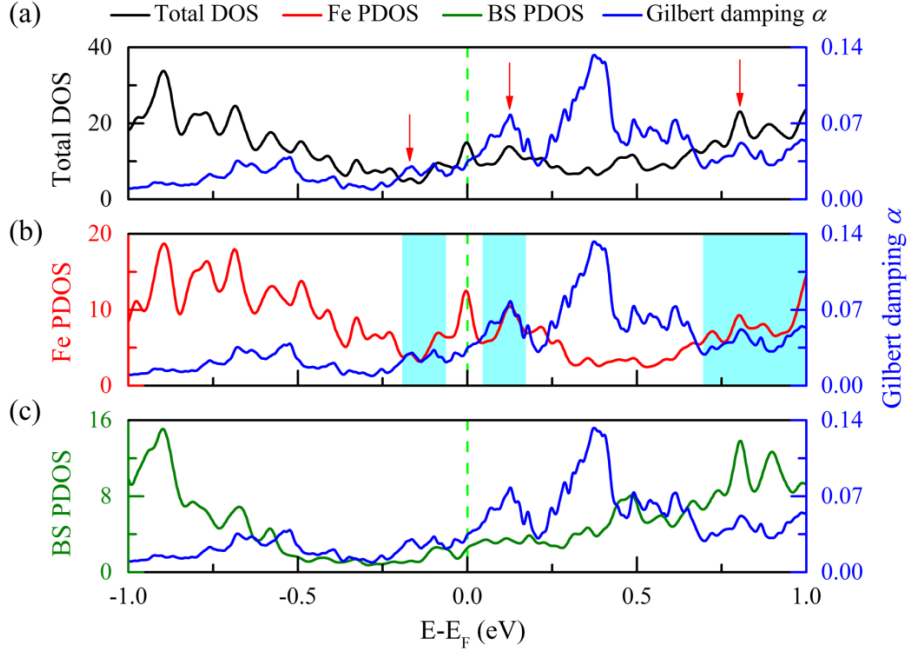

*Figure 5 (color online) Comparisons between Gilbert damping $\alpha$ of BS/Fe at $\Gamma = 26\,\text{meV}$ and (a) total DOS, (b) Fe projected DOS and (c) BS projected DOS. Red arrows and light cyan rectangles highlight the energy windows where Gilbert damping $\alpha$ and the total DOS and Fe PDOS have a strong correlation. In (a), (b) and (c), all DOS are in units of state per eV and Fermi level $E_F$ indicated by the vertical green lines is set to be zero.*

A previous study of Fe, Co and Ni bulks suggested a strong correlation between Gilbert damping and total density of states (DOS) around the Fermi level [36]. To attest if this is applicable here, we show the total DOS and Gilbert damping $\alpha_{\text{BS/Fe}}$ of BS/Fe as a function of the Fermi level based on the rigid band approximation. As shown in Fig. 5a, Gilbert damping $\alpha_{\text{BS/Fe}}$ and the total DOS behave rather differently in most energy regions. From the Fe projected DOS (PDOS) and BS projected PDOS (Fig. 5b and 5c), we see a better correlation between Gilbert damping $\alpha_{\text{BS/Fe}}$ and Fe-projected DOS, especially in regions highlighted by the cyan rectangles. We perceive that although the $\alpha$-DOS correlation might work for simple systems, it doesn't hold when hybridization and SOC are complicated as the effective SOC strength may vary from band to band.

## V. SUMMARY

In summary, we extend our previous torque method from determining magnetocrystalline anisotropy energies [27, 28] to calculating Gilbert damping of FM metals and apply this new approach to Fe, Co and Ni MLs in contact with TI BS. Remarkably, the presence of the TI BS substrate causes order of magnitude enhancements in their Gilbert dampings. Our studies demonstrate such strong enhancement is mainly due to the strong SOC of TI BS substrate. The topological nature of BS may also play a role by facilitating the interfacial hybridization and leaving more states around the Fermi level. Although alloying with heavy elements also enhances Gilbert dampings [32], the use of TIs pushes the enhancement into a much wider range. Our work thus establishes an attractive way for tuning the Gilbert damping of FM metallic films, especially in the ultrathin regime.

## ACKNOWLEDGMENTS

We thank Prof. A. H. MacDonald and Q. Niu at University Texas, Austin, for insightful discussions. We also thank Prof. M. Z. Wu at Colorado State University and Prof. J. Shi at University of California, Riverside for sharing their experimental data before publication. Work was supported by DOE-BES (Grant No. DE-FG02-05ER46237). Density functional theory calculations were performed on parallel computers at NERSC supercomputer centers.

# Appendix A: Details of Gilbert damping calculations

To compare Gilbert dampings of Fe, Co and Ni free-standing MLs with BS/Fe, BS/Co, and BS/Ni, we use $\sqrt{3}\times\sqrt{3}$ supercells containing three atoms and set their lattice constants to 4.164 Å, same as that of the BS substrate. This means that the lattice constant of their primitive unit cell (containing one atom) is fixed at 2.40 Å. The relaxed lattice constants of Fe (2.42 Å), Co (2.35 Å) and Ni (2.36 Å) free-standing MLs are close to this value.

| Systems | $a$ (Å) | $b$ (Å) | $c$ (Å) | k-point grid | $n_{VE}$ | $n_{TB}$ | $\eta$ |
|---|---|---|---|---|---|---|---|
| Fe bulk | 2.931 | 2.931 | 2.931 | 35×35×35 | 16 | 36 | 2.25 |
| Co bulk | 2.491 | 2.491 | 4.044 | 37×37×23 | 18 | 40 | 2.22 |
| Ni bulk | 3.520 | 3.520 | 3.520 | 31×31×31 | 40 | 80 | 2.00 |
| Fe ML | 4.164 | 4.164 | -- | 38×38×1 | 24 | 56 | 2.33 |
| Co ML | 4.164 | 4.164 | -- | 37×37×1 | 27 | 64 | 2.37 |
| Ni ML | 4.164 | 4.164 | -- | 39×39×1 | 30 | 72 | 2.40 |
| BS/Fe | 4.164 | 4.164 | -- | 37×37×1 | 188 | 396 | 2.11 |
| BS/Co | 4.164 | 4.164 | -- | 37×37×1 | 194 | 408 | 2.10 |
| BS/Ni | 4.164 | 4.164 | -- | 37×37×1 | 200 | 432 | 2.16 |
| PbSe/Fe | 4.265 | 4.265 | -- | 37×37×1 | 174 | 376 | 2.16 |

Table A1. Here are details of Gilbert damping calculations of all systems that are studied in this work. $n_{VE}$ is abbreviated for the number of valence electrons and $n_{TB}$ stands for the number of total bands. $\eta$ is the ratio between $n_{VE}$ and $n_{TB}$, namely, $\eta = n_{TB}/n_{VE}$. Note that five QLs of BS are used in calculations for BS/Fe, BS/Co and BS/Ni.

# Appendix B: Derivatives of SOC Hamiltonian $H_{\text{SOC}}$ with respect to the small deviation $u$ of magnetic moments

Based on the SOC Hamiltonian $H_{\text{SOC}}$ in Eq. (5) in the main text, derivatives of the term $\boldsymbol{l}\cdot\boldsymbol{s}$ against the polar angle $\theta$ and azimuth angle $\varphi$ are

$$\frac{\partial}{\partial \theta} \boldsymbol{l} \cdot \boldsymbol{s} = s_n \left( -l_z \sin\theta + \frac{1}{2} l_+ e^{-i\varphi} \cos\theta + \frac{1}{2} l_- e^{i\varphi} \cos\theta \right) +$$
$$\frac{1}{2} s_+ \left( -l_z \cos\theta - \frac{1}{2} l_+ e^{-i\varphi} \sin\theta - \frac{1}{2} l_- e^{i\varphi} \sin\theta \right) + \quad (A1),$$
$$\frac{1}{2} s_- \left( -l_z \cos\theta - \frac{1}{2} l_+ e^{-i\varphi} \sin\theta - \frac{1}{2} l_- e^{i\varphi} \sin\theta \right)$$

and

$$\frac{\partial}{\partial \phi} \boldsymbol{l} \cdot \boldsymbol{s} = s_n \left( 0 + (-i) \times \frac{1}{2} l_+ e^{-i\varphi} \sin\theta + (i) \times \frac{1}{2} l_- e^{i\varphi} \sin\theta \right) +$$
$$\frac{1}{2} s_+ \left( 0 - (-i) \times l_+ e^{-i\varphi} \sin^2 \frac{\theta}{2} + (i) \times l_- e^{i\varphi} \cos^2 \frac{\theta}{2} \right) + \quad (A2).$$
$$\frac{1}{2} s_- \left( 0 + (-i) \times l_+ e^{-i\varphi} \cos^2 \frac{\theta}{2} - (i) \times l_- e^{i\varphi} \sin^2 \frac{\theta}{2} \right)$$

Note that magnetization **M** is assumed to have a constant magnitude when it precesses, so we have $\partial \boldsymbol{H}_{SOC}/\partial M = \partial \boldsymbol{H}_{SOC}/\partial m = 0$. When the normalized magnetization $\boldsymbol{m}$ points along the direction of $\boldsymbol{n}(\theta,\varphi) = (m_x, m_y, m_z)$, we have: $m_x = \sin\theta\cos\varphi$, $m_y = \sin\theta\sin\varphi$ and $m_z = \cos\theta$. Taking $\boldsymbol{m} = \boldsymbol{m}_0 + \boldsymbol{u}$ and the chain rule together, we obtain derivatives of SOC Hamiltonian $\boldsymbol{H}_{SOC}$ with respect to the small deviation $\boldsymbol{u}$ of magnetic moments as follows:

$$\frac{\partial \boldsymbol{H}_{SOC}}{\partial u_x} = \frac{\partial \boldsymbol{H}_{SOC}}{\partial m_x} = \frac{\partial \boldsymbol{H}_{SOC}}{\partial \boldsymbol{m}} \frac{\partial \boldsymbol{m}}{\partial m_x} + \frac{\partial \boldsymbol{H}_{SOC}}{\partial \theta} \frac{\partial \theta}{\partial m_x} + \frac{\partial \boldsymbol{H}_{SOC}}{\partial \varphi} \frac{\partial \varphi}{\partial m_x}$$
$$= \cos\theta\cos\varphi \frac{\partial \boldsymbol{H}_{SOC}}{\partial \theta} - \frac{\sin\varphi}{\sin\theta} \frac{\partial \boldsymbol{H}_{SOC}}{\partial \varphi} \quad (A3),$$

$$\frac{\partial \boldsymbol{H}_{SOC}}{\partial u_y} = \frac{\partial \boldsymbol{H}_{SOC}}{\partial m_y} = \frac{\partial \boldsymbol{H}_{SOC}}{\partial \boldsymbol{m}} \frac{\partial \boldsymbol{m}}{\partial m_y} + \frac{\partial \boldsymbol{H}_{SOC}}{\partial \theta} \frac{\partial \theta}{\partial m_y} + \frac{\partial \boldsymbol{H}_{SOC}}{\partial \varphi} \frac{\partial \varphi}{\partial m_y}$$
$$= \cos\theta\sin\varphi \frac{\partial \boldsymbol{H}_{SOC}}{\partial \theta} + \frac{\cos\varphi}{\sin\theta} \frac{\partial \boldsymbol{H}_{SOC}}{\partial \varphi} \quad (A4),$$

and

$$\frac{\partial \boldsymbol{H}_{SOC}}{\partial u_z} = \frac{\partial \boldsymbol{H}_{SOC}}{\partial m_z} = \frac{\partial \boldsymbol{H}_{SOC}}{\partial \boldsymbol{m}}\frac{\partial \boldsymbol{m}}{\partial m_z} + \frac{\partial \boldsymbol{H}_{SOC}}{\partial \theta}\frac{\partial \theta}{\partial m_z} + \frac{\partial \boldsymbol{H}_{SOC}}{\partial \varphi}\frac{\partial \varphi}{\partial m_z}$$

$$= -\sin\theta \frac{\partial \boldsymbol{H}_{SOC}}{\partial \theta} \qquad (A5).$$

Combining Eq. (5) and Eq. (A1-A6), we can easily obtain the final formulas of derivatives of SOC Hamiltonian $\boldsymbol{H}_{SOC}$ of magnetization $\boldsymbol{m}$.

## Appendix C: Charge transfers and induced magnetic moments in BS/Fe, BS/Co and BS/Ni

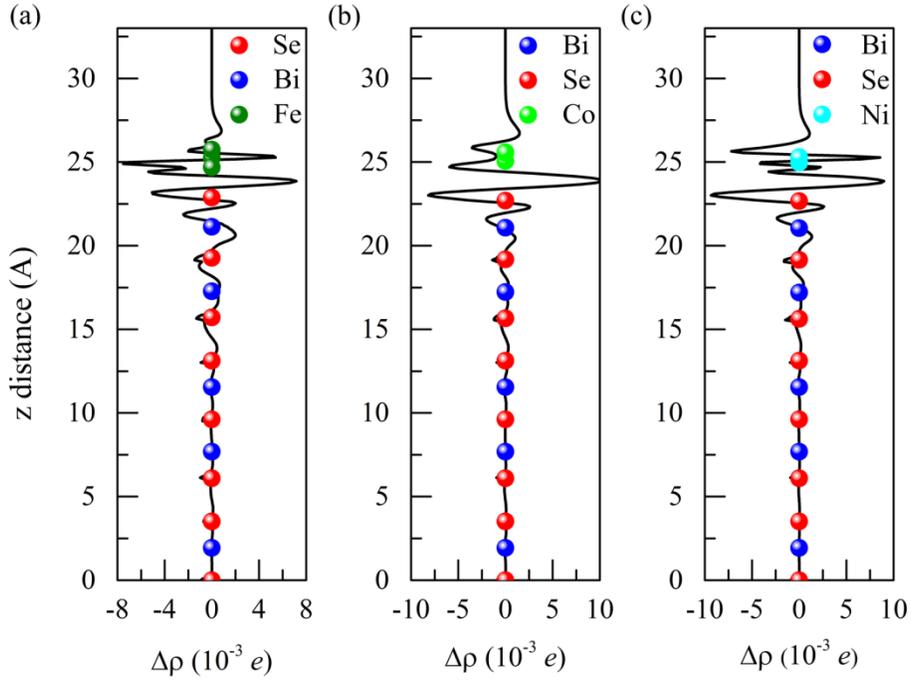

*Figure A1 (color online) Planar-averaged charge difference $\Delta\rho = \rho_{BS+TM-ML} - \rho_{BS} - \rho_{TM-ML}$ (TM = Fe, Co and Ni) of (a) BS/Fe, (b) BS/Co and (c) BS/Ni. The atoms positions are given along the z axis.*

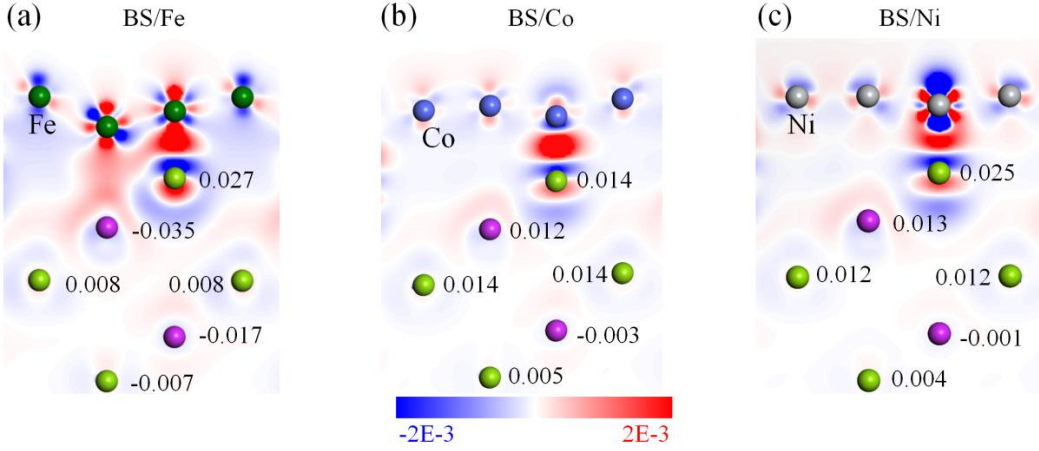

*Figure A2 (Color online) Charge density difference $\Delta\rho = \rho_{BS+TM-ML} - \rho_{BS} - \rho_{TM-ML}$ (TM = Fe, Co and Ni) near the interface between the TM monolayer and the top most QL BS of (a) BS/Fe, (b) BS/Co and (c) BS/Ni. The color bar shows the weights of the negative (blue) and positive (red) charge density differences. Numbers give the induced magnetic moments (in units of $\mu_B$) in the topmost QL BS. Bi and Se atoms are shown by the purple and light green balls, respectively.*

# Appendix D: Contributions of intraband and interband to the Gilbert dampings of BS/Fe, BS/Co and BS/Ni

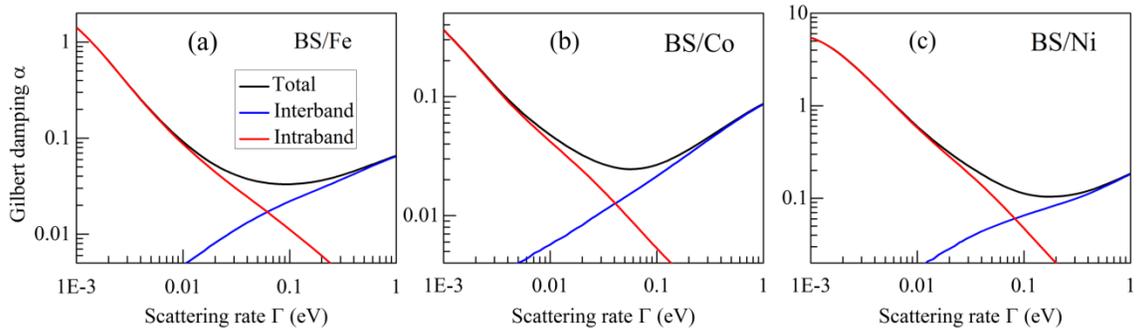

*Figure A3 (color online) Calculated Gilbert dampings of (a) BS/Fe, (b) BS/Co and (c) BS/Ni. Black curves give the total damping. Red and blue curves give the intraband and interband contributions, respectively.*

## Appendix E: Gilbert dampings of BS/Fe with five- and six-QLs of BS slabs

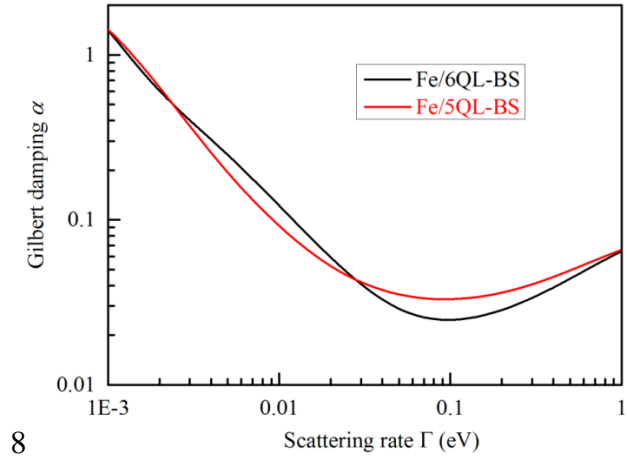

*Figure A4 (color online). Gilbert dampings of BS/Fe with five (red) and six (black) QLs of BS slabs. In the calculations of the Gilbert damping of BS/Fe with six QLs of BS, we use a 39×39×1 Gamma-centered k-point grid, and the number of the total bands is 448 which is twice larger than the number of the total valence electrons (216).*

## Appendix F: Structural comparisons between BS/Fe and PbSe/Fe

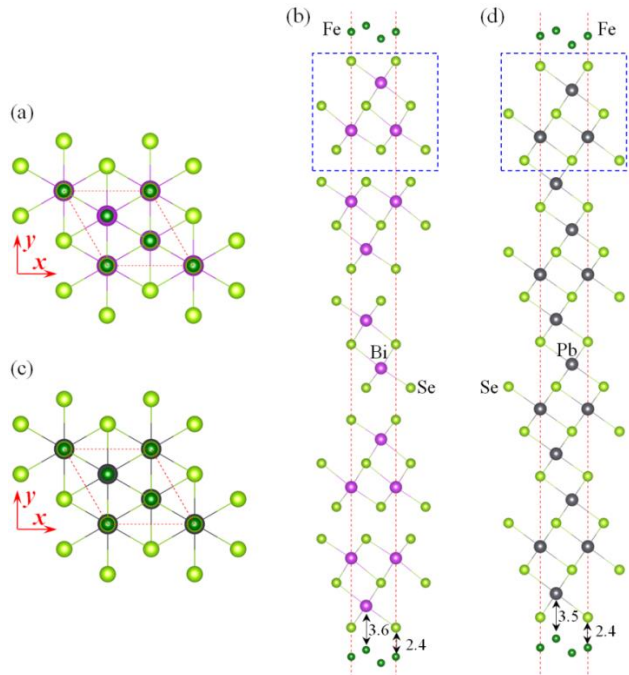

*Figure A5 (color online) (a) Top view and (c) side view of atom arrangement in BS/Fe. (b) Top view and (d) side view of atom arrangement in PbSe/Fe. In (a) and (c), the xyz-coordinates are shown by the red arrows. In (b) and (d), the rectangles with blue dashed lines highlight the most top QL BS in BS/Fe which is similar to the Pb and Se atom layers in PbSe/Fe. The important Fe-Bi, Fe-Se and Fe-Pb bond length is given by the numbers in units of Å. Dark green, light green, purple-red and dark gray balls represent Fe, Se, Bi and Pb atoms, respectively. Note that computational details are given in Table A1.*